\documentclass{cimento_PN}

\usepackage{graphicx}\usepackage{epsfig}
\usepackage{amsmath}\usepackage{amssymb}\usepackage{amsfonts}
\usepackage{bm}
\usepackage{exscale}\usepackage{rotating}
\usepackage{dcolumn}
\usepackage{color}
\usepackage{enumitem}
\usepackage{soul}
\usepackage{ulem}

\title{Is the caloric curve a robust signal of the phase transition in hot nuclei?}
\author{E.~Vient\from{ins1}
\\ For the INDRA and ALADIN Collaborations}
\instlist{\inst{ins1}Normandie Univ, ENSICAEN, UNICAEN, CNRS/IN2P3, LPC Caen, 14000 Caen, France
}

\begin{document}

\maketitle

\begin{abstract}
The richness of the data set, collected by the INDRA collaboration during the last twenty years, enabled us to build a set of caloric curves for nuclei of various sizes, by using, for the first time, a single experimental set-up and a single experimental protocol.  We will therefore present the different caloric curves ($E^{\ast}-T$) obtained by a new calorimetry, for Quasi-Projectiles produced by symmetric or quasi symmetric reactions at different incident energies (Au+Au, Xe+Sn, Ni+Ni). For all these systems, a clear change of the de-excitation process of hot nuclei is observed but this one is neither a plateau nor a back-bending, but a sudden change of slope.
\end{abstract}

\section{Introduction}

Since a long time, the measure of the caloric curves of hot nuclei is a way to study the phase transition of nuclear matter \cite{natowitz1}.
But we may wonder if these caloric curves are really a robust signal of the phase transition in nuclei. We will try to answer this question.
With the different campaigns done by the INDRA collaboration we can build different caloric curves of Quasi-Projectiles for different symmetric or quasi symmetric systems.
The main opportunity for this study is the possibility to use the same multidetector array and a single experimental protocol.
The calorimetry, used here, is a new one, called "3D calorimetry"  \cite{vient1,legouee1, vient2} which allows to determine the excitation energy $E^{\ast}$ of the QP. The temperatures $T$ are estimated from the slopes of kinetic energy spectra defined in the reconstructed Quasi-Projectile frame. We have tried to optimize at the most these two methods of measurement.

\section{Presentation of the necessary event selections}

First, to well experimentally characterize and reconstruct a hot nucleus and specially a QP, we need to do event selections. 
Consequently, we use an event generator here, HIPSE \cite{lacroix1} and a software filter to simulate the experimental response of INDRA. We can thus understand and control our experimental work. 
We verify the correct detection of particles and fragments coming from the QP by the two following conditions : 

\[ 1.05 >\frac{\sum_{i=1}^{Total\,Mul}(ZV_{//})_{i}}{Z_{Proj}\times V_{Proj}} > 0.7 \]
 and \[ 1.05 > \frac{\sum_{i=1}^{Forward\,Mul}(2\times Z_{i})}{Z_{Proj}+Z_{Targ}} > 0.7\]

By this good measurement of the total charge in the forward hemisphere of the center of mass and this correct conservation of the total parallel pseudo-momentum, we obtain a criterion of completeness for the event detection in the forward hemisphere of the center of mass. We can study with HIPSE the consequences of these both selections on the impact parameter distribution. We keep only 28 $\%$ of the total cross section according to HIPSE with these selections for the system  Xe + Sn at 50 A.MeV.

We need also a supplementary selection. Indeed, we must try to control the geometry and the violence of the collision. 
For that, we use the normalized total transverse kinetic energy of Light Charged Particles ($Z<3$), which is define by the following relation :
\[E_{tr12}^{Norm}=\frac{\sum_{i=1}^{Mul_{LCP}} T_{k_i}  \times sin^2(\theta_i)} {(2E_{Available\,in\,c.m.}/3)}\]
Where $T_{k_i}$ is the kinetic energy and $\theta_i$ the polar angle in the laboratory frame.\\
For a completely dissipative collision, this global variable is equal to one.
In reference \cite{vient1} for HIPSE, it is shown a clear mean correlation between this variable and the impact parameter for all the events seen by INDRA. The same one is also observed for the experimental data. For HIPSE, this mean correlation remains the same for the complete events in the forward hemisphere of the center of mass. 

We chose then to select events with this global variable. We divide its distribution in ten slices corresponding to the same cross section.

\section{Fast description of the 3D calorimetry}

\begin{figure}[h]
\centerline{\includegraphics[width=12.46cm,height=3.4cm]{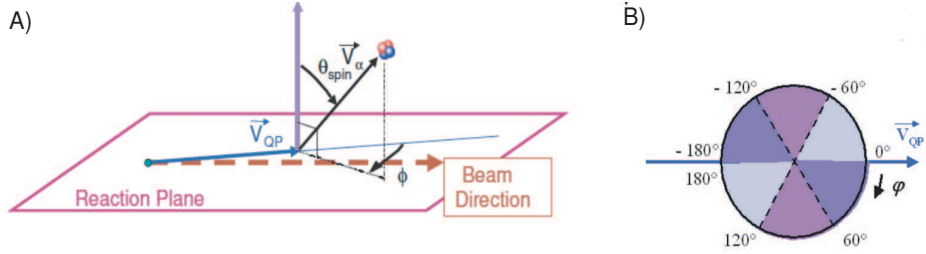}}
\caption{A) Angle definitions in the QP frame. B) Presentation of angular domains of selection.}
\label{fig01} 
\end{figure}

\begin{figure}[h]
\centerline{\includegraphics[width=9.53cm,height=10.14cm]{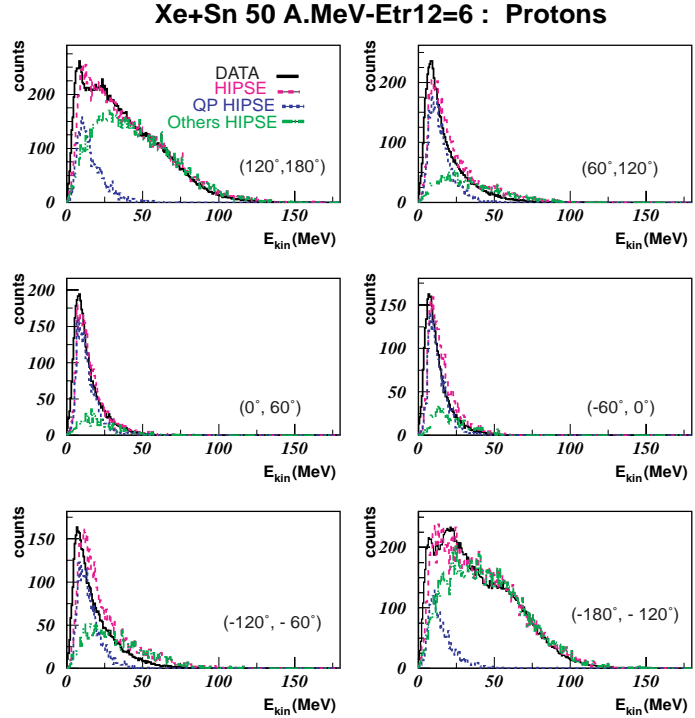}}
\caption{Kinetic energy spectra for the different angular domains obtained for semi-peripheral collisions Xe + Sn at 50 A.MeV. The black curves correspond to the data, the pink to HIPSE, the blue to HIPSE for the protons emitted by the QP and green for the protons with another origin.}
\label{fig02} 
\end{figure}

Now, we will remind quickly what is the 3D calorimetry. The QP frame is reconstructed with IMF's and fragments located in the forward hemisphere of the center of mass. We define the direction of any particle in the QP frame with two angles : the azimuthal angle $\phi$ in the reaction plane, and the polar angle $\theta_ {spin}$, out-of-plane (see part A of figure \ref{fig01}).
In the QP frame, we decided to divide the whole space in six spatial zones with a selection using the azimuthal angle $\phi$ as shown in the part B of figure \ref{fig01}. These spatial domains represent the same solid angle in this frame.
In figure \ref{fig02}, the kinetic energy histograms of protons in the QP frame for the six spatial zones are presented for semi-peripheral collisions, for the system Xe + Sn at 50 A.MeV. The black graphs correspond to the data, the pink to HIPSE. We have simply normalized the two histograms  to the same number of events.  The agreement between the data and HIPSE is really remarkable. With HIPSE,  the origin of the protons is known. Therefore, we know that the blue curves correspond to the protons emitted by the QP and the green curves to the others contributions. We see in this figure that we have a superposition of the blue and pink curves only for the angular zone : $0^{\circ}$, $-60^{\circ}$. We can note that there is a small green contribution also in this zone.

For the 3D calorimetry of the QP, we have chosen to consider that all the particles located in the azimuthal angular range ($0^{\circ}$, $-60^{\circ}$) in the reconstructed frame of the QP are strictly evaporated particles by the QP.
We estimate the evaporation probability according to the kinetic energy and the angular domain by comparison with this reference domain. For example, we divide, for a kind of particles, the kinetic energy distribution of the reference domain by the kinetic energy distribution of another angular domain. We thus obtain an experimental distribution for probability of emission by the QP for this kind of particle.
We can then use these probabilities $Prob_i$ defined for different particles, kinetic energy, $\theta_{spin}$, $\phi$, normalized total transverse kinetic energy of LCP's  to do a calorimetry of the QP, event by event, using different formulas to obtain the charge, the mass and the excitation energy of the QP :
\[E^{\ast}_{QP}=\sum_{i=1}^{Multot} Prob_{i}\times Ec_{i}+ N_{neutron}\times 2\times\langle T \rangle_{p+\alpha}-Q -Ec_{QP} \]
\[Z_{QP}=\sum_{i=1}^{Multot} Prob_{i}\times Z_{i}  \; and \; A_{QP}=Z_{QP} \times 129/54=\sum_{i=1}^{Multot} Prob_{i}\times A_{i}+N_{neutron}\]
\section{Application of the 3D Calorimetry to obtain caloric curves}

\begin{figure}[h]
\centerline{\includegraphics[width=13.69cm,height=5.53cm]{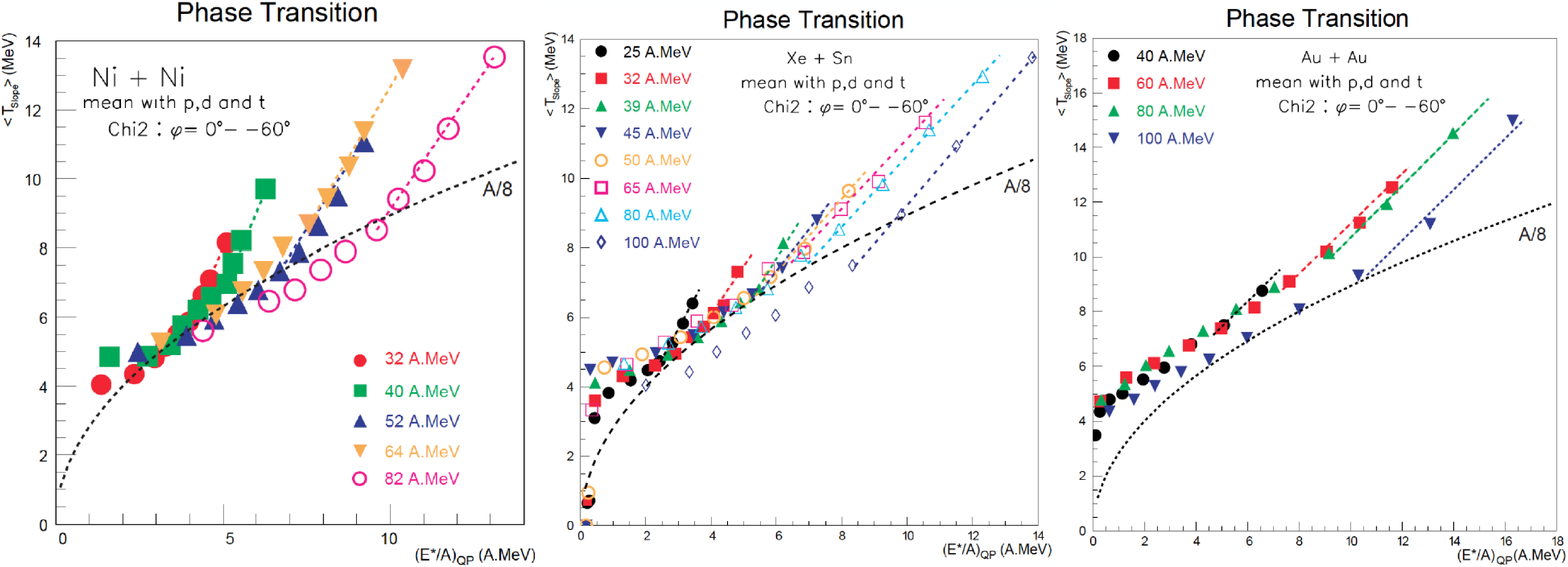}}
\caption{Experimental caloric curves obtained for the systems Ni+Ni, Xe+Sn and Au+Au at different incidental energies.}
\label{fig03} 
\end{figure}

We have applied this 3D calorimetry to different symmetric or quasi symmetric systems : Ni+Ni, Xe+Sn and Au+Au (see figure \ref{fig03}).  The measured temperature is an average temperature calculated using the slopes obtained by a fit of the spectra of protons, deuterons and tritons found for the azimuthal angular domain of reference. 
We can remark two important facts.
First, we see a systematic bump for the peripheral collisions and an apparent drift of the temperature upward  with the size of the system. 
It is a consequence of a complex effect due to the detection and to the criteria of completeness \cite{vient1, vient2}. 
Second, we observe a systematic change of slope for all systems. It seems to correspond to a change of the de-excitation mode of the nuclei. The slope of this change seems to evolve with the size of the system. Does it mean that we have reached the limiting temperature of existence of hot nuclei ? 
In fact, it is difficult to conclude. We know that we have experimental limitations concerning the measure of the temperature.

\begin{figure}[h]
\centerline{\includegraphics[width=9.72cm,height=7.56cm]{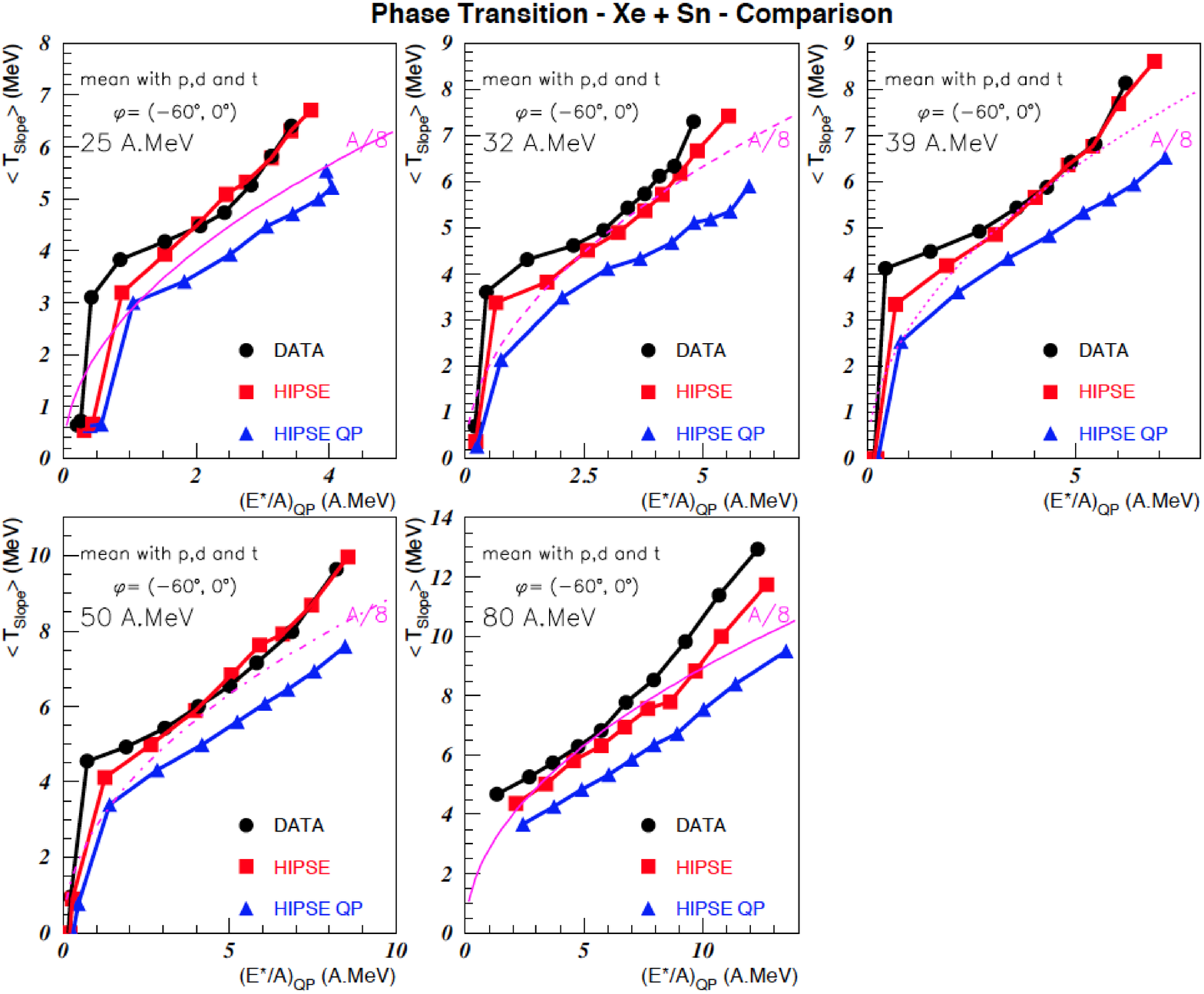}}
\caption{Experimental caloric curves obtained for the system Xe+Sn at different incidental energies for data and HIPSE, with the 3D calorimetry (black circles and red squares respectively) and with a perfect calorimetry (blue triangles).}
\label{fig04} 
\end{figure}

To better understand that, we can observe what happens with HIPSE. We applied the 3D calorimetry to filtered events generated by HIPSE for the system Xe+Sn for different beam energies. The obtained caloric curves are presented in figure \ref{fig04}. In this figure, we compare the data (black circles), HIPSE (red squares) and the data supplied by HIPSE when a perfect calorimetry is applied (blue triangles).
With HIPSE, we known if a nucleus has been or not emitted by the QP. With this information, it is easy to do a perfect calorimetry.
We see clearly that there is an effect due to a pollution by the others contributions in the azimuthal angular domain taken as reference of the QP contribution.
The problem concerning the separation of the different contributions in the nuclear reaction stays experimentally challenging and not completely resolved. It is our main experimental problem. 
\section{Which are the observed signals of phase transition ?}
\begin{figure}[h]
\centerline{\includegraphics[width=13.73cm,height=7.761cm]{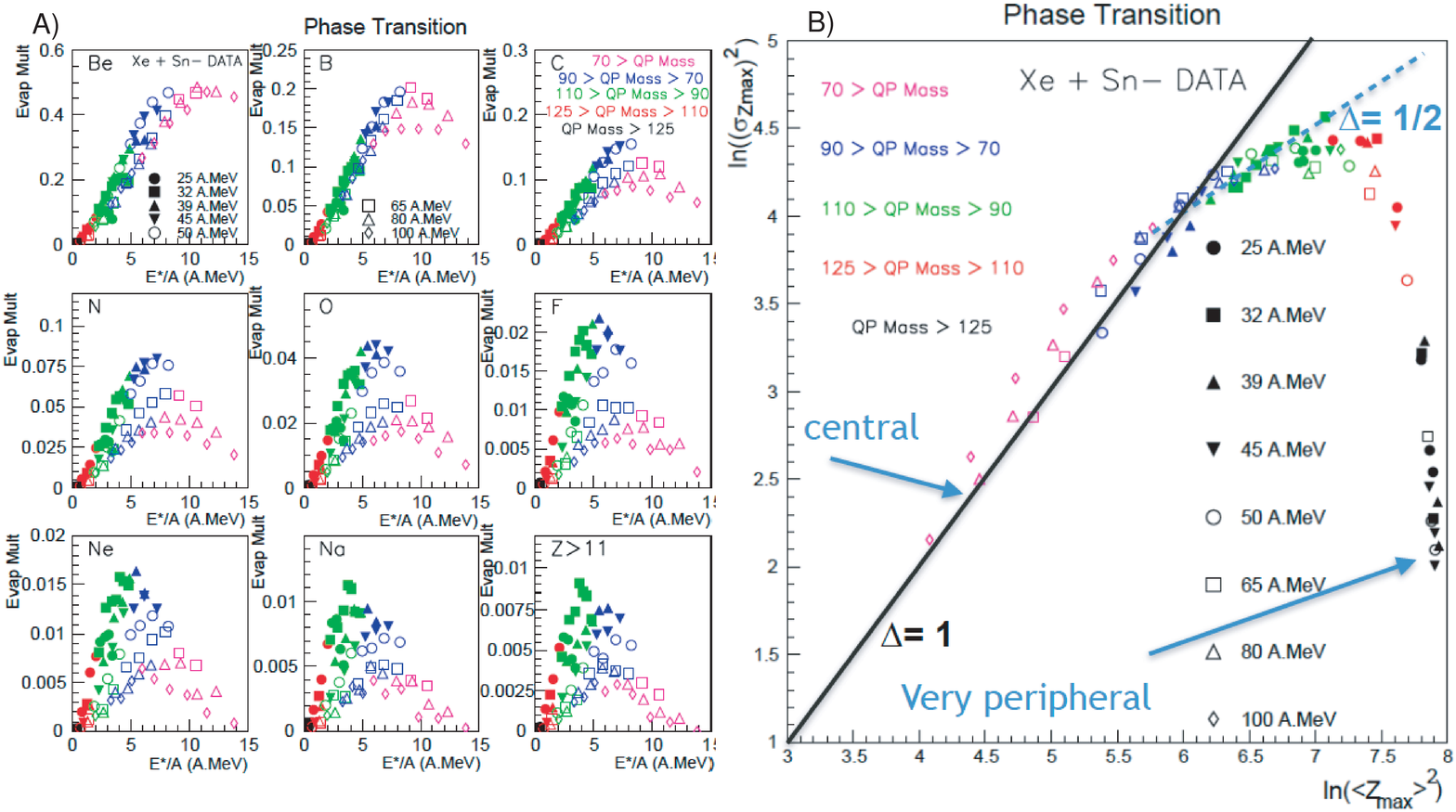}}
\caption{A) Evolution of the mean IMF multiplicity according to the excitation energy per nucleon for the system Xe+Sn at different incident energies. B) Evolution of $ln((\sigma_{Z_{max}})^2)$ as a function  of  $ln(\left\langle Z_{max}\right\rangle^2) $ for the system Xe+Sn at different incident energies.}
\label{fig05} 
\end{figure}

We can then ask: is there a phase transition for the Quasi-Projectiles in these data? For Xe+Sn, if we study the evolution of the mean evaporation multiplicity of different types of particles and fragments of the QP as a function of its excitation energy per nucleon, we observe a rise and fall for these IMF's \cite{kreutz1} (see the part A of the figure \ref{fig05}). On these graphs, the symbol corresponds to a beam energy and the color to a range of QP mass.
We can observe also another possible signal of phase transition : the delta-scaling in the liquid part \cite{frankland1}.
For that, we present in a graph the logarithm of the variance of the charge of the largest fragment in the forward hemisphere as a function of the logarithm of the square of its average value (see the part B of the figure \ref{fig05}). We observe a scaling with $\Delta =1$ for the most central and violent collisions of the higher beam energies as in reference \cite{frankland1}. 
\section{Conclusions and outlooks}

Concerning the question of the existence of a phase transition, the answer is not clear enough for the moment. To get a more definitive answer, we must still improve the QP calorimetry. It means that we must determine the efficiency of the multi-detector array with an event generator or a model sufficiently realistic and later correct its effect.  We must still optimize the separation criteria between the different contributions (QP, QT, pre-equilibrium), ever with an event generator or a model. But these corrections are evidently model-dependent.

\end{document}